\documentclass[preprint]{aastex}
\usepackage{amsmath}
\usepackage{array}
\usepackage{float}
\usepackage{graphicx}
\usepackage{subfigure}

\usepackage{natbib}

\begin{document}
\title{The differing magnitude distributions of the two Jupiter Trojan color populations}
\author{Ian Wong}
\affil{Division of Geological and Planetary Sciences, California Institute of Technology \\
Pasadena, CA 91125, USA}
\email{iwong@caltech.edu}

\and

\author{Michael E. Brown}
\affil{Division of Geological and Planetary Sciences, California Institute of Technology \\
Pasadena, CA 91125, USA}
\and

\author{Joshua P. Emery}
\affil{Department of Earth and Planetary Sciences, University of Tennessee \\
Knoxville, TN 37996, USA}

\begin{abstract}
The Jupiter Trojans are a significant population of minor bodies in the middle Solar System that have garnered substantial interest in recent years. Several spectroscopic studies of these objects have revealed notable bimodalities with respect to near-infrared spectra, infrared albedo, and color, which suggest the existence of two distinct groups among the Trojan population. In this paper, we analyze the magnitude distributions of these two groups, which we refer to as the red and less red color populations. By compiling spectral and photometric data from several previous works, we show that the observed bimodalities are self-consistent and categorize 221 of the 842 Trojans with absolute magnitudes in the range $H<12.3$ into the two color populations. We demonstrate that the magnitude distributions of the two color populations are distinct to a high confidence level ($>95$\%) and fit them individually to a broken power law, with special attention given to evaluating and correcting for incompleteness in the Trojan catalog as well as incompleteness in our categorization of objects. A comparison of the best-fit curves shows that the faint-end power-law slopes are markedly different for the two color populations, which indicates that the red and less red Trojans likely formed in different locations. We propose a few hypotheses for the origin and evolution of the Trojan population based on the analyzed data.
\end{abstract}
\emph{Keywords:} minor planets, asteroids: general

\section{Introduction}

The Jupiter Trojans are a collection of asteroids that lie in a 1:1 mean motion resonance with Jupiter and are confined to two extended swarms centered about the L$_{4}$ and L$_{5}$ Lagrangian points, which lead and trail the planet's motion by an angular distance of $\sim 60$ degrees. Since the first such asteroid was discovered more than a century ago, thousands of Trojans have been confirmed, and the current catalog contains over 6000 objects ranging in size from (624) Hektor, with a diameter of roughly 200 km, to subkilometer-sized objects. Estimates of the total number of Trojans larger than 1 km in diameter range from $\sim 1.0\times10^{5}$ \citep{nakamura} to $\sim2.5\times10^{5}$ \citep{szabo}, corresponding to a bulk mass of approximately $10^{-4}$ Earth masses. These values are comparable with those calculated for main belt asteroids of similar size, making the Trojans a significant population of minor bodies located in the middle Solar System. The orbits of Trojans librate around the stable Lagrangian points with periods on the order of a hundred years and are stable over the age of the Solar System, although long-timescale dynamical interactions with the other outer planets decrease the regions of stability and lead to a gradual diffusion of objects from the Trojan swarms \citep{levisonnature}. Escaped Trojans may serve as an important source of short-period comets and Centaurs, a few of which may have Earth-crossing orbits \citep{marzari}.

Due to their peculiar location and dynamical properties, Trojans lie at the intersection of several of the most important topics in planetary science. The origin and evolution of this population have been a subject of particular interest in recent decades. Early theories proposed a scenario in which the Trojans formed at the same heliocentric distance as Jupiter. In this model, Trojans were created out of the body of planetesimals and dust in the primordial solar nebula that remained after the runaway mass accretion phase of Jupiter and were subsequently stabilized into their current orbits around the Lagrangian points \citep{oldmodel}. However, it has been shown that such \textit{in situ} formation at 5.2~AU cannot explain the presently observed total mass and broad orbital inclination distribution. A recent theory, known as the Nice model, suggests a more complex picture in which the Trojan population originated in a region beyond the primordial orbit of Neptune, and the orbits of Jupiter and Saturn were initially situated much closer to the Sun than they are now \citep{tsiganis}. Through interactions with neighboring planetesimals and perhaps an encounter with a large Neptune-sized body \citep{nesvorny}, the gas giants underwent a rapid migration, crossing resonances and setting off a period of chaotic dynamical alterations in the outer Solar System. It is hypothesized that during this time, the primordial trans-Neptunian planetesimals were disrupted, and a fraction of them were scattered inwards and captured by Jupiter as Trojan asteroids, while the remaining objects were thrown outwards to larger heliocentric distances and eventually formed the Kuiper belt \citep{morbidelli}.

The current understanding of the composition of Trojan asteroids remains incomplete. Visible spectroscopy has shown largely featureless spectra with spectral slopes ranging from neutral to moderately red \citep[e.g.,][]{dotto,fornasier,melita}. Spectroscopic studies of Trojans have also been carried out in the near-infrared, a region that contains absorption bands of materials prevalent in other minor body populations throughout the Solar System, such as hydrous and anhydrous silicates, organics, and water ice \citep[e.g.,][]{emerybrown,dotto,yangjewitt,emery}. These spectra were likewise found to be featureless and did not reveal any incontrovertible  absorption signals to within noise levels. As such, models of the composition and surface properties of Trojans remain poorly constrained. However, several authors have noted bimodality in the distribution of various spectral properties: Bimodality in spectral slope has been detected in both the visible \citep{szabo,roig,melita} and the near-infrared \citep{emery}. The infrared albedo of Trojans has also been shown to display bimodal behavior \citep{grav}. These observations indicate that the Trojans may be comprised of two separate sub-populations that categorically differ in their spectroscopic properties.

While future spectroscopic study promises to improve our knowledge of Trojan composition and structure, a study of the size distribution, or as a proxy, the magnitude distribution, may offer significant insight into the nature of the Trojan population. The magnitude distribution preserves information about the primordial environment in which the Trojans were accreted as well as the processes that have shaped the population since its formation, and can be used to test models of the origin and evolution of the Trojans. In particular, an analysis of the distribution of the attested sub-populations may further our understanding of how these sub-populations arose and how they have changed over time. In this paper, we use published photometric and spectroscopic data to categorize Trojans into two sub-populations and compare their individual magnitude distributions. When constructing the data samples, we evaluate and correct for incompleteness to better model the true Trojan population. In addition to fitting the magnitude distributions and examining their behavior, we explore various interpretations of the data.

\section{Trojan data}
Several sources were consulted in compiling the Trojan data samples analyzed in this work. They are described in the following.

\subsection{Selection of Trojan Dat Samples}
The primary data set is comprised of Trojan asteroids listed by the Minor Planet Center (MPC),\footnote{\texttt{www.minorplanetcenter.org} (Accessed 2014 May 10).} which maintains a compilation of all currently-confirmed Trojans. Absolute magnitude ($H$) and orbital parameter values were taken off of Edward Bowell's ASTORB datafile.\footnote{\texttt{http://www.naic.edu/~nolan/astorb.html} (Accessed 2014 April 27).} The resulting data set, referred to in the following as the \textit{main sample}, contains 6037 Trojans. Of these, 3985 are from the L$_{4}$ swarm and 2052 are from the L$_{5}$ swarm, corresponding to a leading-to-trailing number ratio of 1.94. This significant number asymmetry between the two swarms has been widely noted in the literature and appears to be a real effect that is not attributable to any major selection bias from Trojan surveys, at least in the bright end of the asteroid catalog \citep{szabo}. The brightest object in the main sample has an absolute magnitude of 7.2, while the faintest object has an absolute magnitude of 18.4. The vast majority of Trojans in the main sample (4856 objects) have $H\ge 12.5$, with most of these faint asteroids having been discovered within the last 5 years. In the literature, estimates of the threshold magnitude below which the current total Trojan asteroid catalog is complete lie within the range $H \sim 10.5 - 12$. Therefore, it is only possible to adequately analyze the magnitude distribution of faint Trojans if appropriate scaling techniques are invoked to correct for sample incompleteness. These techniques are discussed in Section 3.2.

Another data set used in this work consists of observations from the fourth release of the Moving Object Catalog of the Sloan Digital Sky Survey (SDSS-MOC4). The SDSS-MOC4 contains photometric measurements of more than 470,000 moving objects from 519 observing runs obtained prior to 2007 March. Of these objects, 557 have been identified to be known Trojans listed in the ASTORB file (243 from L$_{4}$ and 314 from L$_{5}$), and will be referred to in the following as the \textit{Sloan sample}. This data sample includes measured flux densities in the \textit{u, g, r, i, z} bands, centered at 3540, 4770, 6230, 7630, and 9130 {\AA}, respectively, and with bandwidths of $\sim 100$ \AA. As discussed in detail by \cite{szabo}, the distribution of the positions of SDSS observing fields through June 2005 in a coordinate system centered on Jupiter indicates that both L$_{4}$ and L$_{5}$ Trojan swarms were well-covered (i.e., the positions of the observing fields cover a wide range of orbital eccentricity and relative longitude values consistent with Trojan asteroids). Those authors identified 313 known Trojans in the SDSS-MOC3 (previous release) and determined that the survey detected all known Trojans within the coverage area brighter than $H=12.3$. Observing runs since then have expanded the coverage of the sky to include new Trojan swarm regions, yielding 244 additional known Trojans. It is expected that the detection threshold of the Sloan survey (i.e., magnitude to which the SDSS has detected all Trojans within its observing fields) in these newly-covered regions is similar to that determined for the previously-covered regions, and therefore, we may consider our Sloan sample to be a reliable subset of the total Trojan population up to $H\sim 12.3$. This means that the detection threshold of the Sloan sample lies at least 1~mag fainter than the completeness limit of the main sample mentioned above. As part of the analysis presented in the next section, we will confirm the detection threshold of the Sloan sample and use it to arrive at a better estimate of the completeness of the main sample.

\subsection{Categorizing Trojans}
Recent observational studies have identified bimodality in the Trojan population with respect to various photometric and spectroscopic quantities. In this work, we used three earlier analyses of Trojans to classify objects into two color populations.

In \cite{emery}, near-infrared (0.7$-$2.5 $\mu$m) spectra of 58 Trojans were collected during four observing runs at the NASA Infrared Telescope Facility and were combined with previously-published spectra of 10 other Trojans. Together, these objects range in magnitude from $H = 7.2$ to $H = 10.7$. For each object, the authors measured the reflectance fluxes in four bands, centered at 0.85, 1.22 (J-band), 1.63 (H-band), and 2.19 (K-band) $\mu$m, from which color indices were calculated using $m_{\lambda1}-m_{\lambda2} = 2.5\log(R_{\lambda2}/R_{\lambda1})$, where $m_{\lambda1}-m_{\lambda2}$ is the color index for two wavelengths, and $R_{\lambda2}/R_{\lambda1}$ is the ratio between the corresponding reflectance fluxes. These color indices quantify the spectral slopes of the Trojans in the near-infrared, with higher index values corresponding to redder spectra. Notably, the plot of the J-K color index versus the 0.85-J color index for the asteroids analyzed is not continuous; rather, there is a distinct break separating a redder group (Group I) from a less red group (Group II). The distribution of the 0.85-H color index likewise shows a clear bimodality, while the H-K histogram is unimodal, suggesting that the difference between the two groups of asteroids is concentrated primarily in the short-wavelength end of the near-infrared spectrum ($\lambda < 1.5$~$\mu$m). Both L$_{4}$ and L$_{5}$ swarms were shown to display similar bimodal behavior, and it was determined that the two identified groups in the analyzed Trojan sample could not have been drawn from a unimodal distribution to a very high confidence level ($>99.99\%$). We included the color indices of 15 additional Trojans (Emery et al., in preparation) for a total of 83 objects, which we categorized into Group I (19 objects) and Group II (64 objects).

\cite{grav} presented thermal model fits for 478 Trojans observed with the Wide-field Infrared Survey Explorer (WISE), which conducted a full-sky survey in four infrared wavelengths: 3.4, 4.6, 12, and 22 $\mu$m (denoted W1, W2, W3, and W4, respectively). Using the survey data, the W1 albedo was computed for each object, and it was shown that the distribution of W1 albedos as a function of diameter is discernibly bimodal for the 66 objects with diameters larger than $\sim$60 km, which corresponds to objects brighter than $H\sim 9.6$; for the smaller (fainter) Trojans, the errors in the measured albedos are much larger, and a clear bimodality was not discernible. Among these 66 large Trojans, 51 have W1 albedo values between 0.11 and 0.18 (Group A), while 15 have W1 albedo values between 0.05 and 0.10 (Group B). Within each group, the albedo values show no dependence on diameter and are tightly clustered, with average separations between adjacent albedo values of 0.001 and 0.004 for Group A and Group B, respectively. 

More importantly, when considering the Trojans that are in both the \cite{grav} and the \cite{emery} data sets, one finds that every object in Group A is a member of Group I, and every object in Group B is a member of Group II, with the sole exception of (1404) Ajax, which has high H-K and 0.85-J color indices characteristic of redder Group I objects, but a relatively low W1 albedo value of 0.085. This correspondence between groups categorized with respect to different spectroscopic quantities reinforces the proposal presented by \cite{emery} that the Trojans are comprised of two distinct populations with dissimilar spectral properties and likely different compositions. In particular, we conclude that Group I and Group A are both sampled from one of the two Trojan populations; these objects have redder color indices, and we will refer to this population as the red (R) population. Analogously, Group II and Group B are both sampled from the second Trojan population, which will be referred to as the less red (LR) population, due to the relatively lower near-infrared color indices of its members.

Using the robust and consistent bimodalities observed by \cite{emery} and \cite{grav}, we categorized 93 Trojans as either LR (20 objects) or R (73 objects). However, these population sizes are too small to allow for statistically meaningful statements about the overall Trojan population. Moreover, the faintest object in this group has an absolute magnitude of $H=10.7$, which would restrict our analysis of the Trojan color populations to just the relatively bright objects. In order to expand our categorization of Trojans into color populations, we turned to photometric data from the Sloan survey.

\cite{roig} studied 250 known Trojans from the SDSS-MOC3 and computed spectral slopes from the listed \textit{u, g, r, i, z} band flux densities. The authors noted that the distribution of spectral slopes is bimodal. We expanded on this study, reproducing the spectral slope calculations and including new Trojans listed in the SDSS-MOC4. Following the procedure used in \cite{roig}, we corrected the flux densities using the solar colors provided in \cite{ivezic}:  $c_{u-r}=(u-r)-1.77$, $c_{g-r} = (g-r)-0.45$, $c_{r-i}=(r-i)-0.10$, and $c_{r-z}=(r-z)-0.14$. The reflectance fluxes, $F$, normalized to $1$ in the \textit{r} band, were defined as: $F_{u} = 10^{-0.4c_{u-r}}$, $F_{g}=10^{-0.4c_{g-r}}$, $F_{i} =10^{0.4c_{r-i}}$, and $F_{z} = 10^{0.4c_{r-z}}$. The relative errors $\Delta F/F$ were estimated using the second-order approach in \cite{roiggilhutton}:
\begin{equation}\label{error}\Delta F/F = 0.9210\Delta c(1+0.4605\Delta c),\end{equation}
where the color errors $\Delta c$ are computed as the root-squared sum of the corresponding magnitude errors, e.g., $\Delta c_{u-r} = \sqrt{(\Delta u)^{2}+(\Delta r)^{2}}$. The error in $F_{r}$ was estimated using $\Delta c_{r-r} = \sqrt{2}\Delta r$. We discarded all asteroid observations that had a relative error greater than 10\% in any of the fluxes besides $F_{u}$, which usually has larger errors due to the effects of instrument noise in and around the \textit{u}-band. We also considered only asteroids with magnitudes in the range $H<12.3$, over which the Sloan survey is expected to have detected all Trojans within its survey area.

\begin{figure}[H]
\begin{center}
\includegraphics[width=9cm]{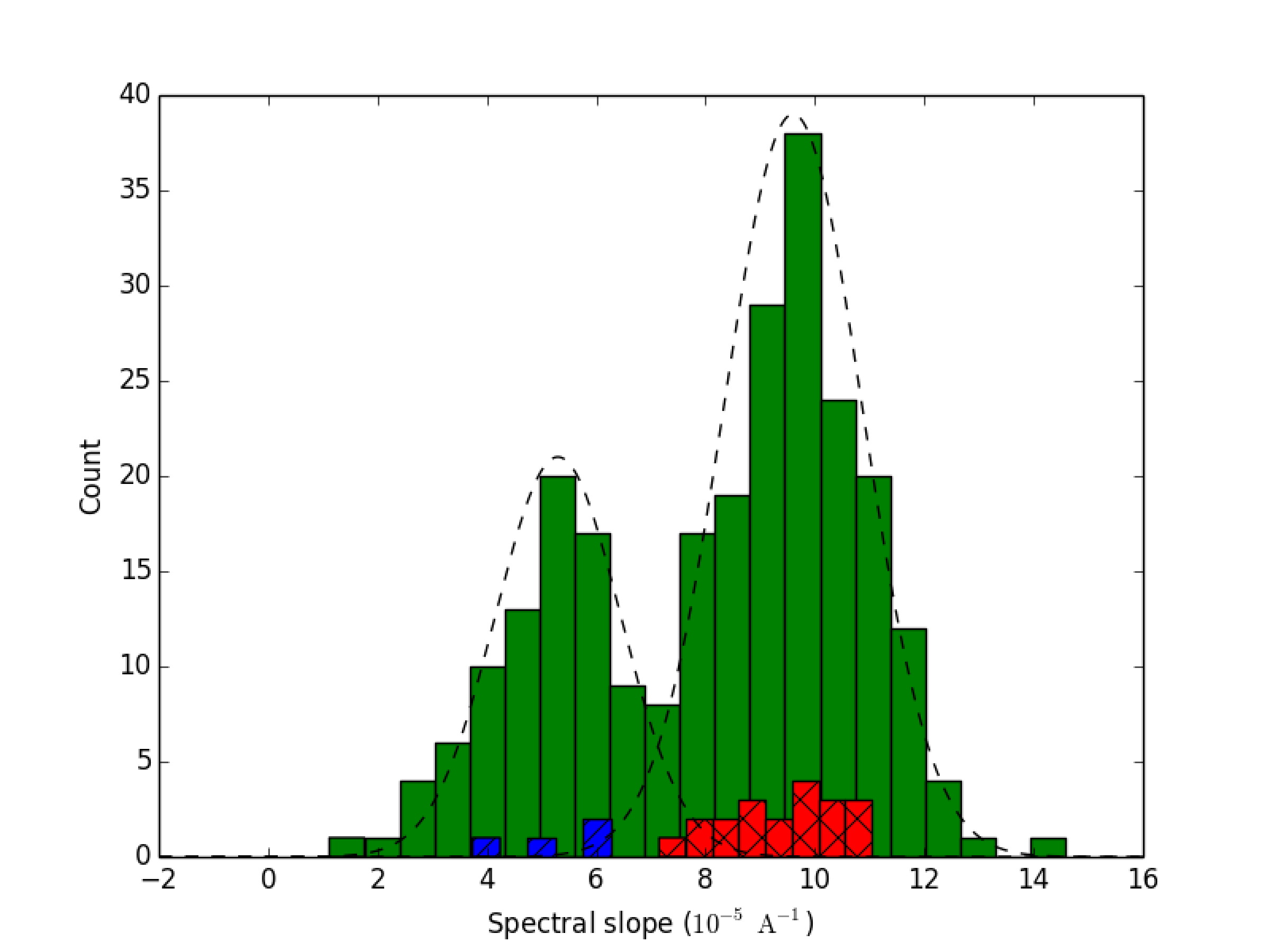}
\end{center}
\caption{Distribution of spectral slopes of all 254 Trojans in the Sloan sample with $H< 12.3$ (solid green), and the distributions of spectral slopes of 24 Trojans classified into the LR and R populations per \cite{emery} and \cite{grav} (blue with diagonal hatching and red with cross hatching, respectively). The best-fit Gaussian distribution functions for the two color populations are shown as black dashed lines.} \label{spectralslope}
\end{figure}

The resulting asteroid set contains 254 objects (114 in L$_{4}$ and 140 in L$_{5}$), 24 of which were included in the \cite{emery} and/or \cite{grav} analyses and previously categorized by spectrum. For each object, the spectral slope $S$ was computed from a linear least-squares fit to a straight line passing through the fluxes $F_{g}$, $F_{r}$, $F_{i}$, and $F_{z}$, taking into account the individual errors $\Delta F$ ($F_{u}$ was not used in this computation, as per \cite{roig}). If an object had multiple observations, the average of the spectral slopes computed for all observations was used. The histogram of spectral slopes is shown in Figure~\ref{spectralslope}. From the plot, the bimodality in the spectral slope distribution is evident.\footnote{In \cite{roig}, it was reported that only objects in the L$_{4}$ swarm showed this bimodality in spectral slope. Our present analysis includes many more asteroids from the SDSS-MOC4, and we observe bimodality in both L$_{4}$ and L$_{5}$ swarms.} By fitting the spectral slope distribution with two Gaussians, we found that one of the two modes is centered at $S =  5.3 \times 10^{-5}$~\AA$^{-1}$, while the other mode is located at higher spectral slopes (i.e., redder colors), with a peak at $S = 9.6 \times 10^{-5}$~\AA$^{-1}$; the best-fit Gaussian distribution functions are plotted in Figure~\ref{spectralslope}. This two-peaked distribution shape is similar to the one presented by \cite{emery} for the H-K color index. In particular, the 24 Trojans in the Sloan sample that have already been categorized into LR and R populations (4 in LR and 20 in R) align with the two modes shown in Figure~\ref{spectralslope}. Therefore, we can say that objects with spectral slope values consistent with the left mode belong to the LR population, while objects with spectral slope values consistent with the right mode belong to the R population. There is some overlap between the two modes, which makes it difficult to categorize all of the Trojans observed by the SDSS into populations. Nevertheless, we may expand our categorization by adopting conservative break-off spectral slope values: All Trojans with $S\le 5.3 \times 10^{-5}$~\AA$^{-1}$ were classified as less red, while all Trojans with $S\ge 9.6 \times 10^{-5}$~\AA$^{-1}$ were classified as red. Using this method, we were able to categorize 151 of the 254 asteroids in the SDSS-MOC4 with $H< 12.3$; 47 objects belong to the LR population, and 104 objects belong to the R population, with the remaining 103 objects being uncategorized.

The estimated 95\% detection flux density thresholds for the \textit{u, g, r, i, z} bands are  22.0, 22.2, 22.2, 21.3, and 20.5, respectively \citep{ivezic}. The average relative band magnitudes for the 151 Trojans in the color populations that were imaged by the SDSS are $u-r = 2.08$, $g-r = 0.62$, $i-r=-0.26$, $z-r = -0.42$ for R objects and $u-r = 2.01$, $g-r = 0.52$, $i-r=-0.18$, $z-r = -0.26$ for LR objects. For an object to be listed on the Moving Object Catalog, it must have detections in at least three bands. The detection threshold in the \textit{z}- and \textit{i}-bands are the lowest. For objects with the same \textit{r}-band magnitude, LR objects are less reflective at longer wavelengths, so for objects with magnitudes near the detection thresholds, there is a bias against LR objects. However, the differences between the relative band magnitudes among the two color populations are not large, and this bias is only expected to affect the objects with absolute magnitudes at the very faint end of our considered range and beyond. Therefore, for our data samples, this effect is minor and is not taken into consideration in our analysis.

\begin{figure}[H]
\begin{center}
\includegraphics[width=9cm]{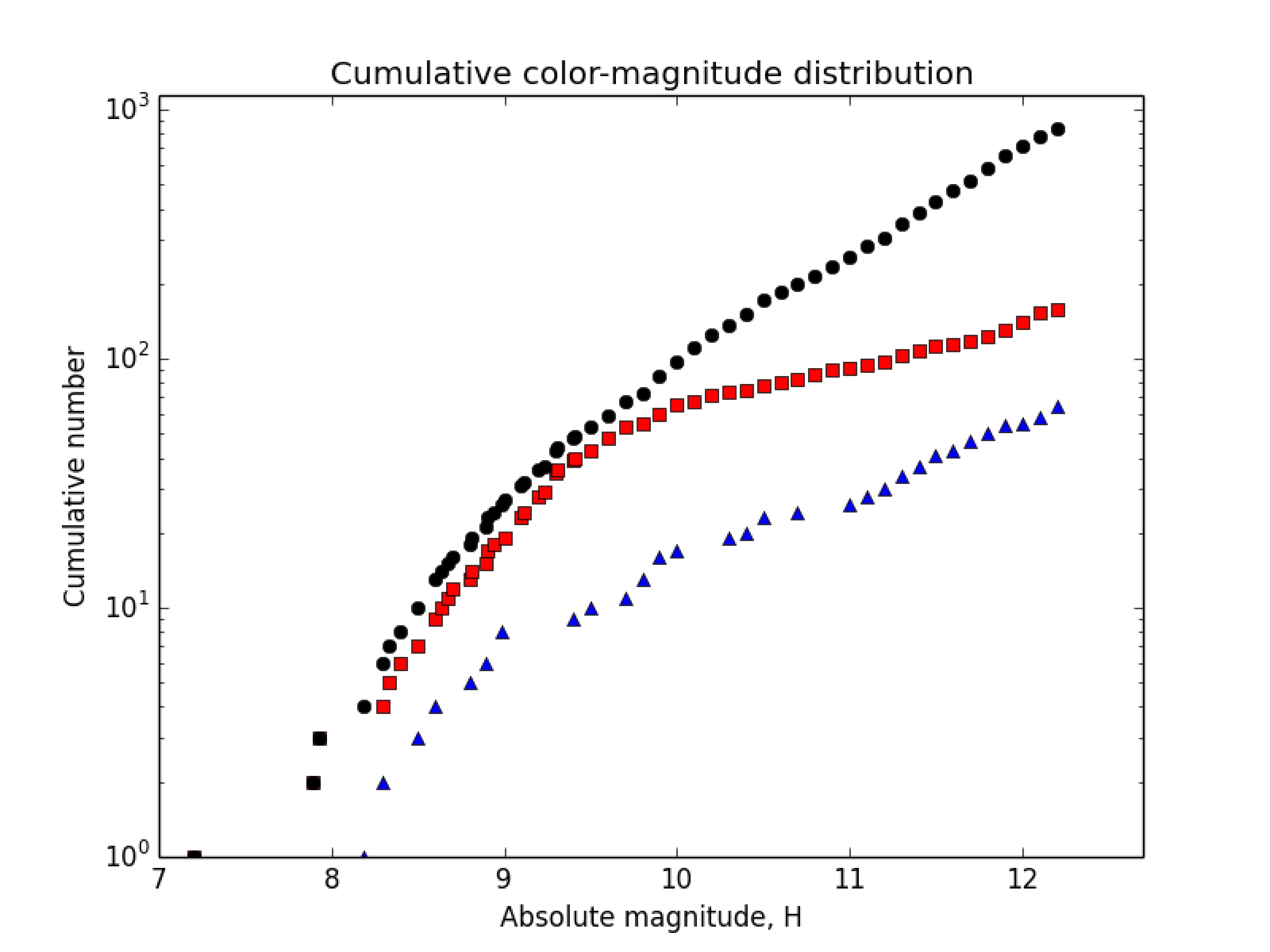}
\end{center}
\caption{Plot of the unscaled cumulative magnitude distributions for the main Trojan sample (black circles) and the categorized R and LR color populations (red squares and blue triangles, respectively). These data have not yet been corrected for incompleteness.} \label{colors}
\end{figure}

We have compared three photometric and spectroscopic studies of Trojans and determined that the bimodal behaviors observed in all these studies are consistent and indicative of the existence of two separate color populations. Of the 842 objects in the main sample with $H< 12.3$, 478 are in the L$_{4}$ swarm, and 364 are in the L$_{5}$ swarm, which entails a leading-to-trailing number ratio of 1.31. This ratio is notably smaller than the value of 1.94 obtained for the total Trojan catalog, which suggests that there may be major detection biases favoring L$_{4}$ Trojans among the faintest objects. After categorizing the objects in the main sample, we found that 64 objects belong to the LR population, and 157 objects belong to the R population, while the remaining 621 objects were not categorized because they have either not been analyzed by any of the three studies discussed above or have spectral slope values between $5.3 \times 10^{-5}$~\AA$^{-1}$ and $9.6 \times 10^{-5}$~\AA$^{-1}$. In Figure~\ref{colors}, the cumulative magnitude distribution $N(H)$, i.e., the total number of asteroids with absolute magnitude less than or equal to $H$, is plotted for the main sample and the two color populations. The distributions plotted here have not been scaled to correct for incompleteness.

\section{Analysis}

In this section, the magnitude distributions of the Trojan populations are studied. We present best-fit curves to describe the magnitude distributions and compare their behavior.

\subsection{Population Distinctness}
Previously, we classified Trojans into LR and R populations based on various spectroscopic and photometric quantities. While the observed bimodalies indicate that the two populations differ categorically with respect to several spectral properties, the current lack of understanding of Trojan surface composition makes it difficult to use these spectral properties in studying the origin and evolution of Trojans. Moreover, the distinction in spectral properties does not preclude the possibility that the Trojans are simply a mixed population of LR and R objects, with a constant number ratio between the two populations at each magnitude. To determine whether the two color populations are distinct, we must compare the shape of their distributions.

While the LR and R populations are incomplete, there is no reason to believe that one of the two populations is significantly more complete than the other. In particular, the ratio of R to LR objects at each magnitude is not expected to be affected by any major bias (see Section~3.2 for details of our analysis of sample completeness). Since the difference in shape of two magnitude distributions is determined largely by the variation of the number ratio of the two distributions with respect to magnitude, we may test for population distinctness of the Trojan color samples by using the current LR and R populations as plotted in Figure~\ref{colors}, without the need to scale up both populations to correct for incompleteness.

Already from the unscaled cumulative magnitude distributions plotted in Figure~\ref{colors}, one can see that the distributions of the color populations are dissimilar. To analytically examine the distinctness of the LR and R populations, we used the two-sample Kuiper variant of the Kolmogorov$-$Smirnov test \citep[Kuiper$-$KS test;][]{press}. This nonparametric statistic quantifies the likelihood that two data samples are drawn from the same underlying distribution. It evaluates the sum of the maximum distances of one distribution above and below the other and returns a test decision value, $p$, between 0 and 1, which represents the probability that the two data samples are not drawn from the same underlying distribution. The Kuiper$-$KS test is sensitive to differences in both the relative location and the shape of the two cumulative distributions. It is particularly appropriate when dealing with distributions that differ primarily in their tails, as is the case with the Trojan color populations. 

Running the Kuiper$-$KS test on the two color populations, we obtained a $p$-value of 0.973. This high test decision value demonstrates that the two color populations are not sampled from a single underlying distribution to a confidence level of 97.3\%. In other words, the LR and R Trojan populations are distinct not only with respect to the spectral properties of their members, but also with respect to their overall size/magnitude distributions.

\subsection{Sample Completeness}
When analyzing a population distribution, it is important to determine and properly correct for any incompleteness in the data sample. To ensure that our curve-fitting adequately models the true Trojan magnitude distribution, we used the Sloan sample to estimate the incompleteness of the main sample and color populations.

As discussed in Section~2.1, the detection threshold of the SDSS within its coverage area is much fainter than the completeness threshold of the overall Trojan catalog. The Sloan survey broadly sampled the orbital parameter space characteristic of both Trojan swarms. Important to our analysis is whether there exists any variation in the magnitude distribution of objects across different regions of the Trojan swarms, since such variation would lead to the total magnitude distribution of the Sloan sample being significantly different from the true total magnitude distribution. Recent studies of Trojans have not observed any discernible correlation between absolute magnitude and eccentricity or inclination in either the leading or the trailing swarm \citep[c.f.,][]{szabo,fernandez}, so it is unlikely that the Sloan sample is characterized by any bias with respect to magnitude. We may therefore consider the Sloan sample to be an accurate scaled-down representation of the overall Trojan population. With the exception of a few bright Trojans, all objects in the data samples have absolute magnitudes given with tenth-place accuracy (e.g., $H = 10.1$); in other words, they are effectively binned into 0.1~mag groups. To evaluate the completeness of our main sample, we examine the ratio $R$ between the cumulative number of objects in the Sloan sample and the cumulative number of objects in the main sample for each 0.1~mag bin. Over the range of magnitudes for which both the main sample and the Sloan sample are complete, $R$ should be roughly constant at some value.  As the magnitude increases up to the detection threshold of the Sloan sample, the main sample becomes incomplete and $R$ should increase steadily. At higher magnitudes, past the detection threshold of the Sloan sample, $R$ is expected to decrease, since a large number of faint Trojans have been discovered since the release of the SDSS-MOC4. 

\begin{figure}[H]
\begin{center}
\includegraphics[width=9cm]{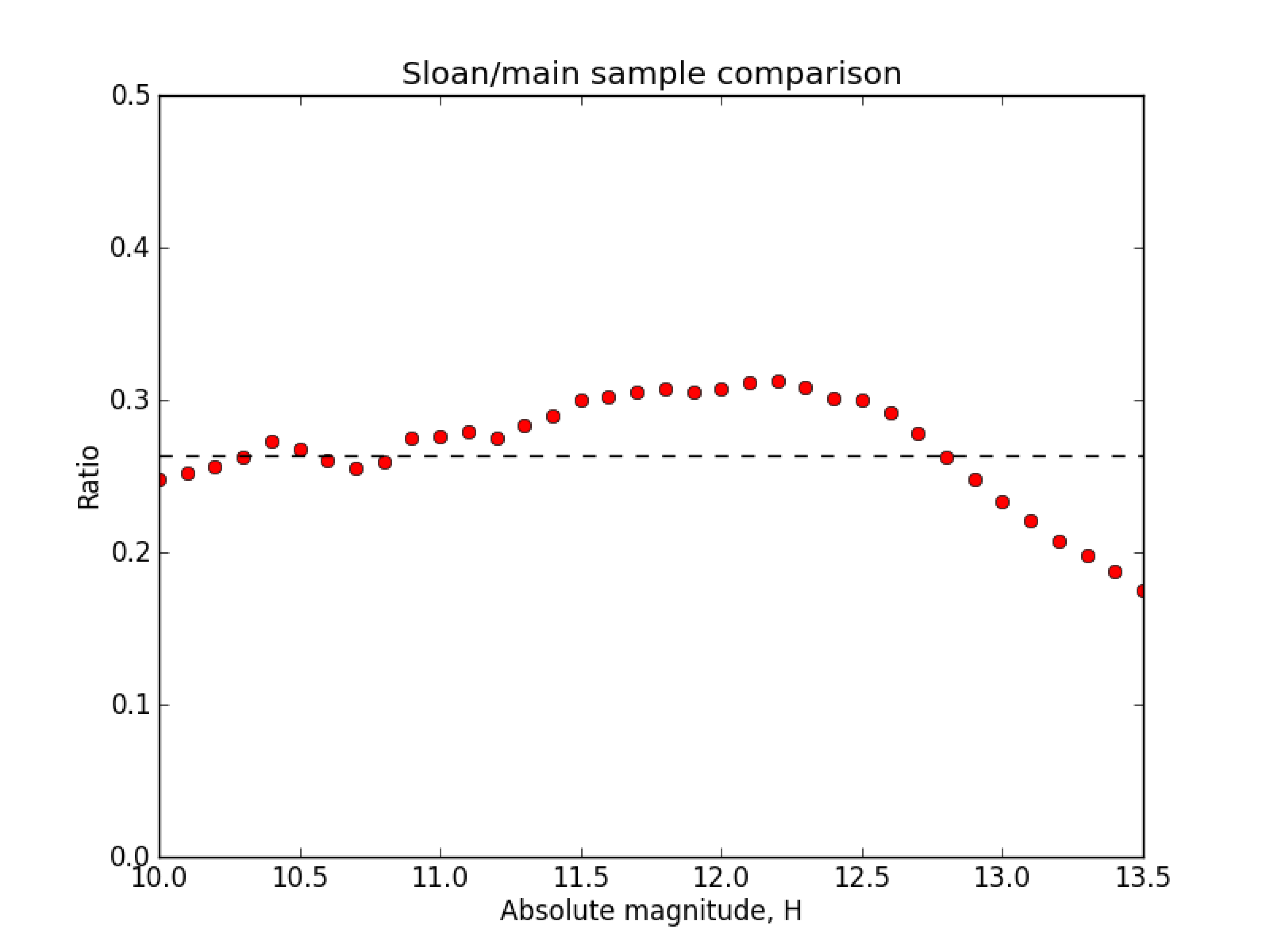}
\end{center}
\caption{Plot of the ratio between the cumulative number of objects in the Sloan sample and the cumulative number of objects in the main sample for various absolute magnitude bins (red circles). The black dashed line indicates the average value $R_{*}$ for bins with $H=10.0 \rightarrow 11.2$.} \label{comparison}
\end{figure}

Figure~\ref{comparison} shows the values of $R$ plotted with absolute magnitude. From the plot, the expected behavior described earlier is evident:  For bins with $H< 11.3$, the value of $R$ is roughly constant at $R_{*} = 0.264$, which is the average of $R$ for bins with $H=10.0 \rightarrow 11.2$. (Bright objects were omitted from the average, since the small bin numbers lead to significant scatter in $R$.) At fainter magnitudes, $R$ increases until $H = 12.3$, after which it decreases rapidly. From this, we conclude that every Trojan brighter than $H=11.3$ is contained in the main sample (that is, the total Trojan catalog), while the Sloan sample is complete up to $H=12.3$ (that is, contains an unbiased subsample of Trojans), which confirms the completeness limit estimate given in \cite{szabo}. Using the calculated values of $R$, we can now evaluate the catalog efficiency $\eta_{\mathrm{mpc}}$ of the main sample, i.e., the ratio of the number of Trojans $n_{\mathrm{mpc}}$ currently cataloged by the Minor Planet Center to the true number of Trojans $n_{0}$, in each bin with $H<12.3$. For $H<11.3$, the main sample is complete, so $\eta_{\mathrm{mpc}}$ = 1. For $11.3\le H < 12.3$, we first evaluate the ratio $r(H)$ between the non-cumulative (i.e., differential or bin-only) number of Trojans in the Sloan and main samples for each 0.1~mag bin; the values of $r(H)$ in this interval are greater than the benchmark value of $r_{*} = \bar{r} = 0.29$, where $\bar{r}$ is the average of $r(H)$ over the interval $H=10.0 \rightarrow 11.2$. The catalog efficiency value for each bin is given by $r(H)/r_{*}$. We subsequently fit a fifth-order polynomial through the binned catalog efficiency values over the domain $11.3\le H < 12.3$ to arrive at a smooth functional form $\eta_{1}(H)$. The catalog efficiency can be expressed as a single piecewise-defined function:
\begin{equation}\label{detection}\eta_{\mathrm{mpc}}(H) = 
\begin{cases}
1,&\text{for }H<11.3\\
\eta_{1}(H),&\text{for }11.3\le H < 12.3
\end{cases}.\end{equation}

More careful consideration must be made when correcting for incompleteness in the color populations. While the absolute magnitude distribution of Trojans does not appear be dependent on the location in orbital parameter space and would not be affected by the particular locations of observed fields within the Trojan swarms, as discussed earlier, correlations between the color of objects and orbital parameters may lead to biases in the resulting magnitude distributions of the color populations. Most Trojans (621 out of 842) were not categorized as either less red or red, with the brightest unclassified asteroid having $H=9.6$. The majority of objects in our color populations (151 out of 221) were classified using the spectral slope categorization method based off Sloan data. Using data from the SDSS-MOC3, \cite{szabo} and \cite{roig} reported a weak correlation between spectral slope and inclination, with objects at larger inclinations tending to be redder; \cite{fornasier} reported a similar correlation in their study of visible spectral slope and interpreted it as a lack of faint objects with low spectral slope. This color-inclination correlation was found to be the same in both swarms. \cite{szabo} identified a bias in their data: the L$_{5}$ subsample of Trojans had a significantly larger fraction of objects with high inclinations than the L$_{4}$ subsample. In our analysis, such asymmetric coverage would cause the number ratio of R-to-LR L$_{5}$ Trojans to be unrealistically inflated and skew the overall color distributions. 

To determine whether a similar bias is present among the 254 objects in the current Sloan sample, we computed the fraction of objects in the SDSS-MOC4 with large inclinations ($i>20^{\circ}$) for the leading and trailing swarms independently. It was found that the fraction is similar for the two swarms (0.24 for L$_{4}$ and 0.22 for L$_{5}$). This means that observing runs since the release of the SDSS-MOC3 have captured more high-inclination regions of the L$_{4}$ swarm, and as a result, the leading and trailing swarms are equally well-sampled in the SDSS-MOC4 data. Therefore, no selection bias with respect to inclination is discernible in the Sloan sample, and we may consider the LR and R color populations defined in Section 2.1 to be a representative subset of the true color composition of the overall Trojan population. In particular, the number ratio of red to less red Trojans in each  bin should be approximately the same as the true ratio at that magnitude. We define a categorization efficiency value for each bin, which is the ratio between the number of already-categorized Trojans in the LR and R populations, $n_{LR}(H)+n_{R}(H)$, and the total number of detected Trojans, $n_{\mathrm{det}}(H)$. Over the domain $9.6\le H < 12.3$, where the color classification is incomplete, we followed a similar procedure to that used in deriving the detection efficiency and fitted a polynomial through the categorization efficiency values to obtain a smooth function $\eta_{2}(H)$. We can write the overall categorization efficiency function as
\begin{equation}\label{categorization}\eta_{\mathrm{cat}}(H) = 
\begin{cases}
1,&\text{for }H<9.6\\
\eta_{2}(H),&\text{for }9.6\le H < 12.3
\end{cases}.\end{equation}
This categorization efficiency function is the same for both LR and R populations and must be coupled with the detection efficiency function $\eta_{\mathrm{det}}(H)$ for $H\ge 11.3$. 

The total efficiency functions for the main sample and color populations, which take into account catalog and/or categorization incompleteness, are given by:
\begin{equation}\label{eta}\eta(H) = 
\begin{cases}
\eta_{\mathrm{mpc}}(H),&\text{for the main sample}\\
\eta_{\mathrm{cat}}(H)\times\eta_{\mathrm{mpc}}(H),&\text{for the LR and R populations}
\end{cases}.\end{equation}
We used catalog and categorization efficiency to scale up the data samples so that they approximate the true Trojan population. Similar scaling methods have been employed in the study of the size distribution and taxonomy of main belt asteroids \citep{demeo}. We demonstrate our method with the following example: at $H=11.5$, there are 43 objects in the main sample, 17 of which are also contained in the Sloan sample. The ratio between the number of objects in the Sloan and main samples is $r = 17/43 \approx 0.395$, which yields a catalog efficiency value of $\eta = r_{*}/r \approx 0.73$. Thus, the approximate true number of Trojans with $H=11.5$ is $n_{0}=43/\eta \sim 59$. The scaled and unscaled cumulative magnitude distributions for the main sample and color populations are shown in Figures~\ref{total}-\ref{LR}.

\subsection{Distribution Fits}

Previous analyses of the magnitude distributions of Trojans (see, for example, \cite{jewitt}) have shown that the differential magnitude distribution, $\Sigma(H) = dN(H)/dH$, is well-described by a broken power law with four parameters:
\begin{equation}\label{distribution}\Sigma(\alpha_{1},\alpha_{2},H_{0},H_{b}|H) = 
\begin{cases}
10^{\alpha_{1}(H-H_{0})},&\text{for }H< H_{b}\\
10^{\alpha_{2}H+(\alpha_{1}-\alpha_{2})H_{b} - \alpha_{1}H_{0}},&\text{for }H \ge H_{b}
\end{cases},\end{equation}
where there is a sudden change from a bright-end slope $\alpha_{1}$ to a shallower faint-end slope $\alpha_{2}$ at some break magnitude $H_{b}$. $H_{0}$ is the threshold magnitude for which $\Sigma(H_{0}) = 1$ and serves to properly normalize the distribution to fit the data. \cite{jewitt} obtained the slope values $\alpha_{1}=1.1$  and $\alpha_{2}=0.4$ from their study of 257 Trojans, which did not correct for incompleteness in the faint-end distribution. More recent studies of faint Trojans by \cite{szabo} and \cite{yoshida} obtained faint-end slope values of 0.44 and 0.38, respectively. 

We fitted the magnitude distributions of the total Trojan sample and the two color populations to the broken power law distribution function in equation~\eqref{distribution} by using a maximum likelihood method similar to the one used in \cite{fraser} for their study of Kuiper belt objects (KBOs). Given a list of Trojan magnitudes and a particular set of parameters for the distribution function to be fitted, this technique defines a likelihood function $L$, which returns the probability that a random sampling of the distribution will yield the data. The maximum likelihood method is well-suited for analyzing data sets like the ones under consideration, since it is robust to  small data counts and non-Gaussian statistics, for which typical $\chi^{2}$ fitting methods are inappropriate. Also, other statistical considerations like catalog and categorization efficiency can be easily integrated into the formulation.
	
The likelihood function used in our fitting takes the form
\begin{equation}\label{likelihood}L(\alpha_{1},\alpha_{2},H_{0},H_{b}| H_{i})\propto e^{-N}\prod_{i}P_{i},\end{equation}
where $H_{i}$ is the absolute magnitude of each detected Trojan, $N$ is the total number of detected objects expected in the magnitude range under consideration, and $P_{i}$ is the probability of having object $i$ with magnitude $H_{i}$ given the underlying distribution function $\Sigma$. Taking into account detection and categorization incompleteness, $N$ is given by
\begin{equation}\label{N}N = \int^{H_{\mathrm{max}}}_{-\infty} \eta(H)\Sigma(\alpha_{1},\alpha_{2},H_{0},H_{b}|H) \,dH,\end{equation}
where $\eta(H)$ is the efficiency function defined in equation~\eqref{eta}, and $H_{max}=12.3$. By including the efficiency function, we ensure that the curves are fitted to the true Trojan distribution, not the incomplete detected Trojan distribution. The probability $P_{i}$ is simply the differential density function evaluated at $H_{i}$, i.e., $P_{i} = \Sigma(\alpha_{1},\alpha_{2},H_{0},H_{b}|H_{i}) $.

The best-fit distribution functions were obtained by maximizing the likelihood function over the four-dimensional parameter space using an affine-invariant Markov chain Monte Carlo (MCMC) Ensemble sampler with 100,000 steps \citep{mcmc}. The optimal parameters and corresponding 1$\sigma$ errors were computed for each distribution. The magnitude distribution of the main sample (all Trojans with $H< 12.3$) is best-fit by $\alpha_{1}=1.11\pm0.02$, $\alpha_{2}=0.46\pm 0.01$, $H_{0}=7.09^{+0.03}_{-0.02}$, and $H_{b}=8.16^{+0.03}_{-0.04}$. The bright-end slope is consistent with the value calculated in \cite{jewitt}, while the faint-slope is steeper than previously-obtained values, due to our correction for incompleteness in the Trojan catalog past $H=11.3$. The L$_{4}$ and L$_{5}$ Trojans were independently analyzed for detection completeness and fitted in a similar fashion. The optimal values of the slopes $\alpha_{1}$ and $\alpha_{2}$ for the two swarm distributions were found to be indistinguishable within calculated uncertainties. This agrees with the results of earlier studies \citep[see, for example,][]{yoshida2} and demonstrates that the leading and trailing Trojan swarms have magnitude distributions that are identical in shape, differing only in overall asteroid number. 

\begin{figure}[H]
\begin{center}
\includegraphics[width=9cm]{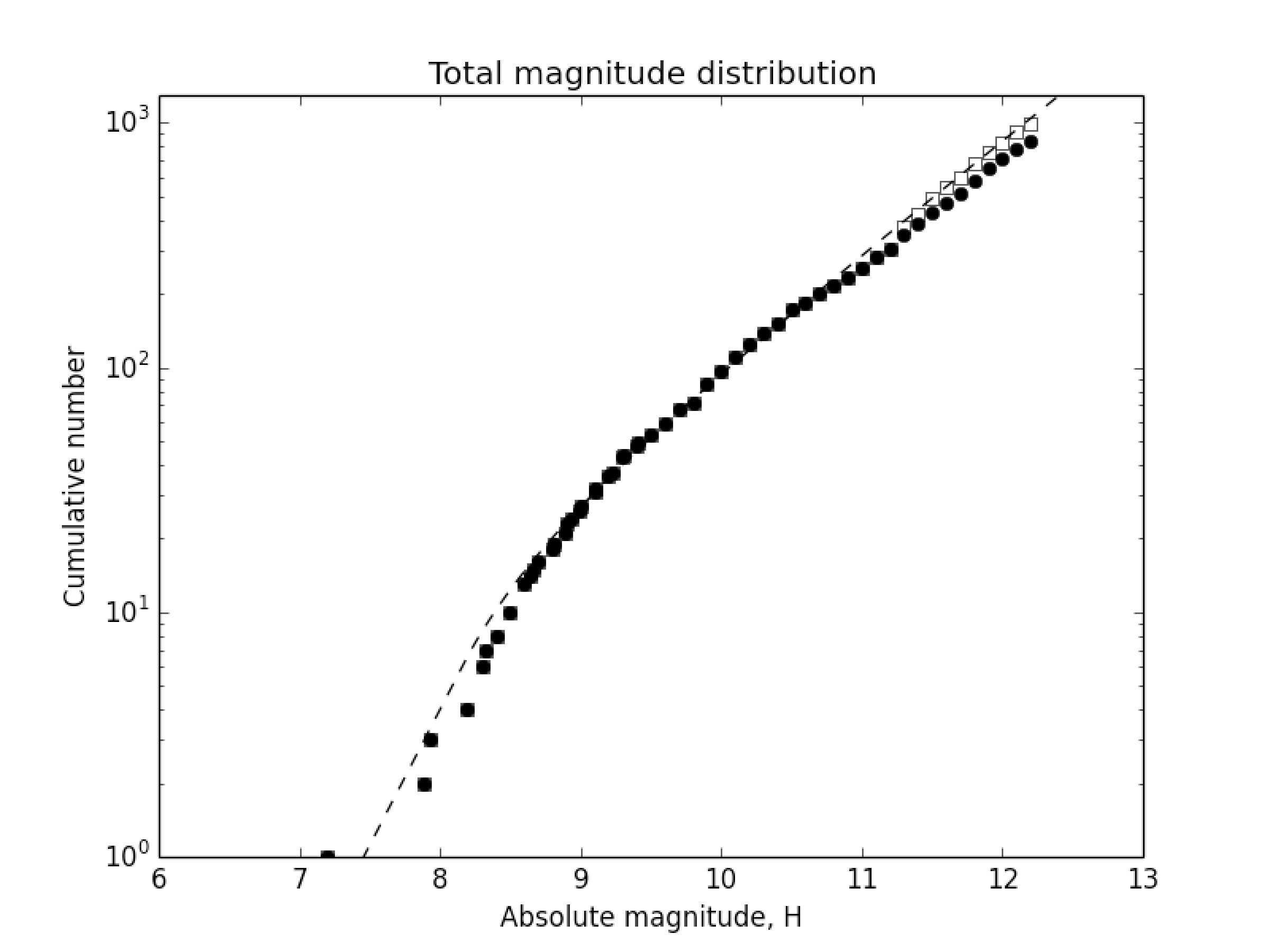}
\end{center}
\caption{Plot depicting the scaled (white squares) and unscaled (black circles) cumulative magnitude distributions for the total Trojan population, along with the best-fit curve describing the true Trojan cumulative distribution.} \label{total}
\end{figure}

\begin{figure}[H]
\begin{center}
\includegraphics[width=9cm]{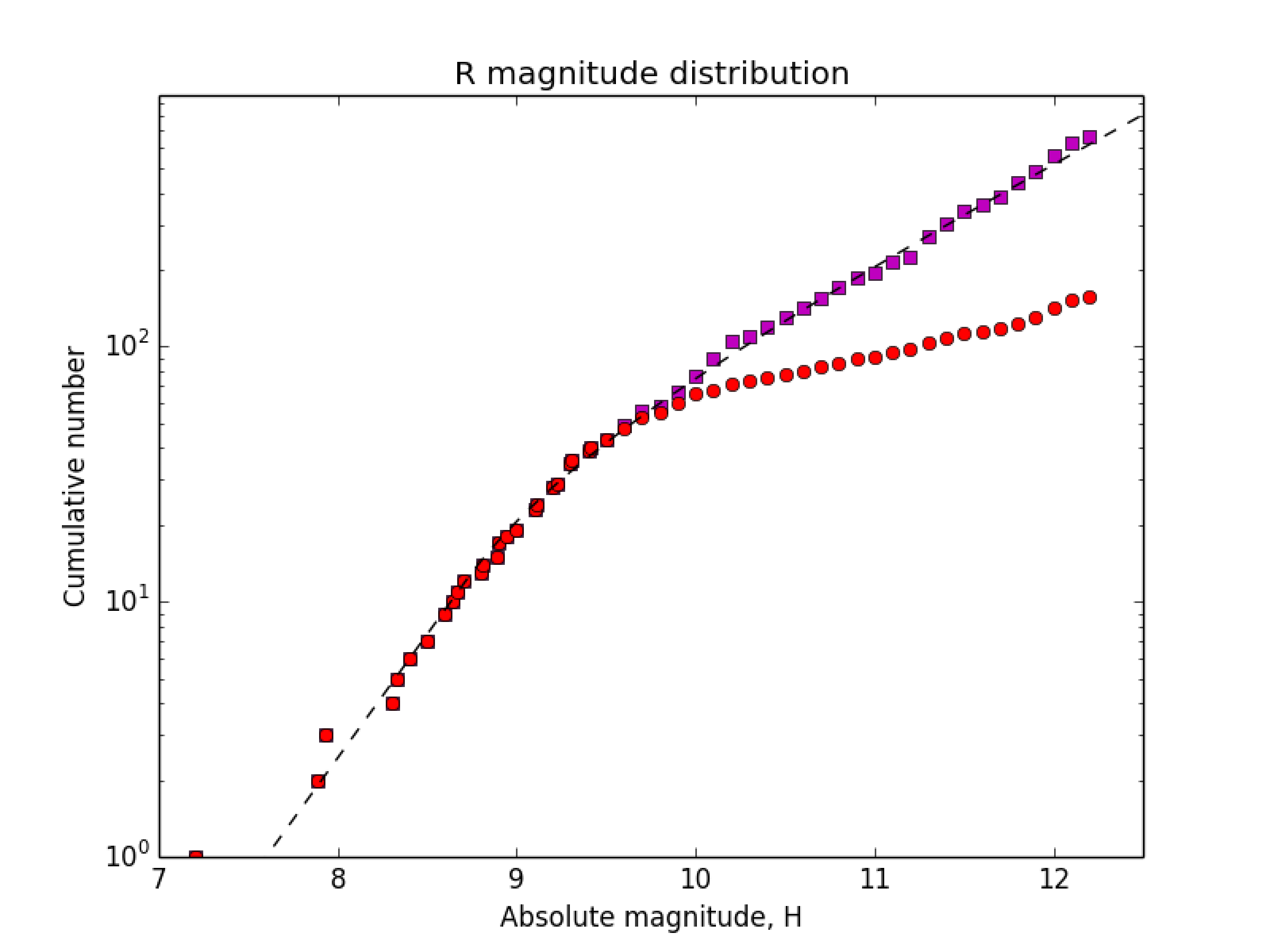}
\end{center}
\caption{Plot depicting the scaled (magenta squares) and unscaled (red circles) cumulative magnitude distributions for R population, along with the best-fit curve describing the true cumulative distribution.} \label{R}
\end{figure}

\begin{figure}[H]
\begin{center}
\includegraphics[width=9cm]{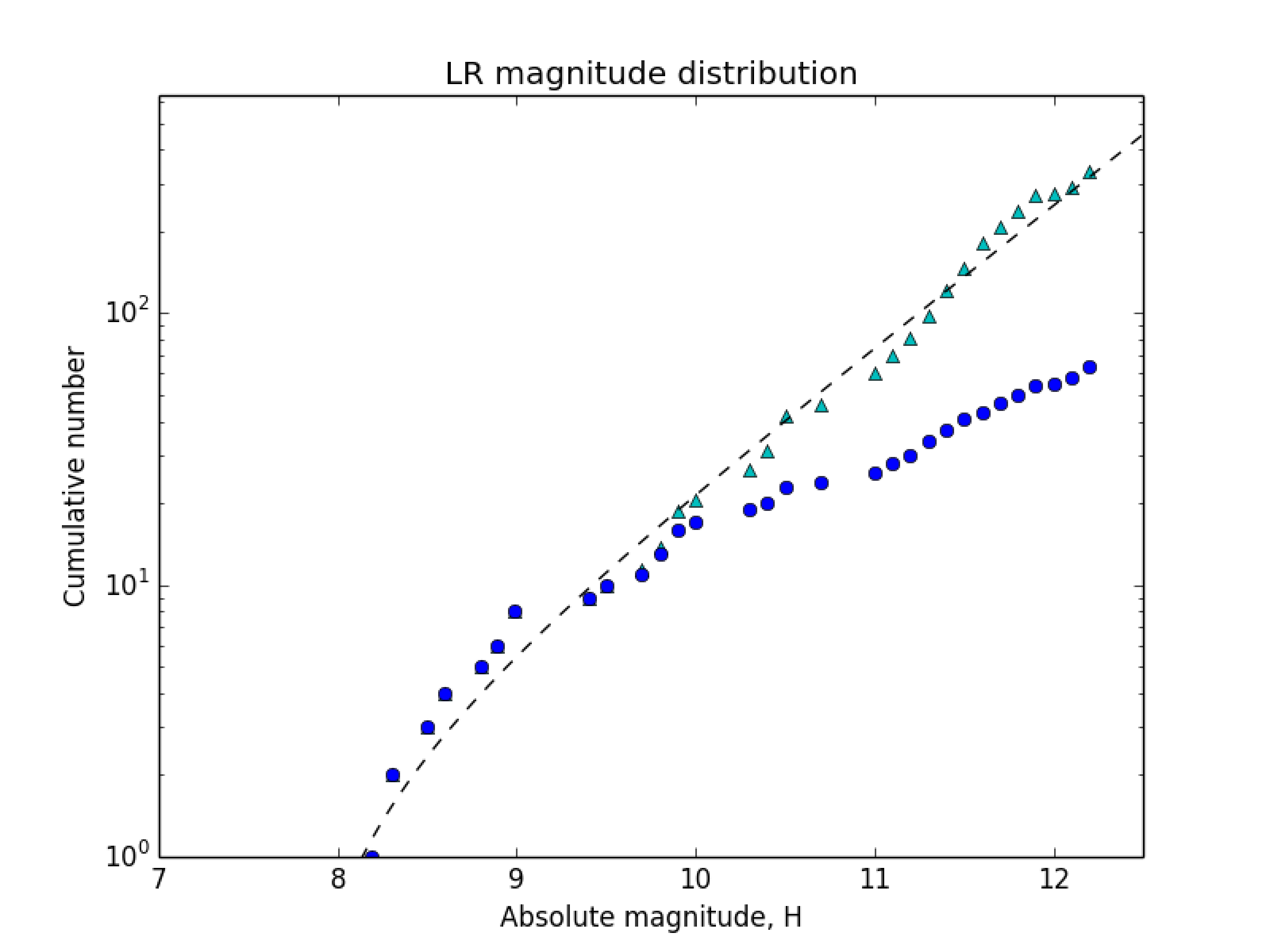}
\end{center}
\caption{Plot depicting the scaled (cyan triangles) and unscaled (blue circles) cumulative magnitude distributions for LR population, along with the best-fit curve describing the true cumulative distribution.} \label{LR}
\end{figure}

The magnitude distributions of the color populations were both individually  fitted to a broken power law. The optimal parameters for the R population magnitude distribution are $\alpha_{1}=0.97^{+0.05}_{-0.04}$, $\alpha_{2}=0.38\pm0.02$, $H_{0}=7.24^{+0.05}_{-0.07}$, and $H_{b}=8.70^{+0.08}_{-0.11}$, while for the LR population magnitude distribution, they are $\alpha_{1}=1.25^{+0.09}_{-0.04}$, $\alpha_{2}=0.52^{+0.03}_{-0.01}$, $H_{0}=7.77^{+0.04}_{-0.09}$, and $H_{b}=8.15^{+0.06}_{-0.10}$. Figures~\ref{total}-\ref{LR} show the cumulative magnitude distributions for the main sample and the color populations, along with the best-fit curves that describe the true distributions. In each plot, the lower distribution is the cumulative count for the unscaled data set, and the upper distribution is the approximate true distribution, scaled to correct for catalog and/or categorization incompleteness (as described in Section 3.2).

\section{Discussion}

The analysis of the Trojan magnitude distributions in the previous section, which utilized the most current asteroid catalog and corrected for catalog incompleteness, presents the most accurate picture to date of the true Trojan population up to the cutoff magnitude $H=12.3$.
Using the distribution fits calculated for both the total and the two color populations, we can try to reach a better understanding of the origin and evolution of the Trojans, and in particular, the nature of the two color populations.

The most notable feature of the magnitude distributions is the transition from a steep power-law slope to a shallower slope at $H\sim8 - 9$. Previous studies of the total Trojan magnitude distribution \citep[e.g.,][]{morbidelli} have suggested that the broken power-law shape separates the population into two groups: Objects with magnitudes brighter than the break magnitude are described by a power-law slope that reflects the primordial accretion processes that created the original Trojan population. On the other hand, objects with magnitudes fainter than the break magnitude form a sub-population that has reached collisional equilibrium and is mostly comprised of collisional fragments of larger objects. It was demonstrated in pioneering work by \cite{dohnanyi} that the magnitude distribution of a small body population that evolves solely through self-collisions attains an equilibrium power-law slope of $\alpha_{*} \sim 0.5$ when collisional equilibrium is achieved, regardless of the initial shape of the distribution. The faint-end slope of the total Trojan magnitude distribution that we obtained by fitting the data is $\alpha_{2} = 0.46\pm0.01$, which is consistent with the canonical collisional equilibrium slope. In relation to the history of the Trojan population, there arises the question of whether the sharp roll-over to a shallower faint-end slope in the currently-observed population is a consequence of collisional evolution after the Trojans were emplaced in their current orbits around Jupiter, or a result of collisional interactions in the primordial trans-Neptunian region prior to emplacement. Several authors have modeled the collisional evolution of Trojans and determined that the observed broken power-law distribution is best reproduced when assuming that a break was present at the time of emplacement \citep[c.f.,][] {marzari,deelia}. Furthermore, these studies have shown that the intrinsic collision probabilities characteristic of the Trojan swarms are insufficiently high to have brought about any significant collisional evolution among objects with magnitudes brighter than the break. Thus, the currently-observed bright-end distribution reflects the shape of the primordial size distribution of large Trojans at the time of emplacement.

A more peculiar aspect of the Trojans is the magnitude distributions of the color populations - in particular, the difference between the faint-end slopes of the R and LR populations ($0.38\pm0.02$ and $0.52^{+0.03}_{-0.01}$, respectively). The Kuiper$-$KS test demonstrated that the magnitude distributions of the color populations are remarkably distinct, which indicates that the two populations likely formed in different places before being emplaced into the Trojan regions. While the fitted bright-end slopes are different, the distinction is most apparent in the faint-end portion of the distributions. (Running the Kuiper$-$KS test on just the bright-end portions of the color distributions yielded intermediate $p$-values, which are inconclusive as a metric for population distinctness.) A hypothesis that posits a scenario in which the two color populations arose from different regions in the primordial trans-Neptunian disk would be able to explain the different bright-end slopes, which are determined primarily by the accretion environment. However, in light of the interpretation that the faint-end portion of the broken power-law distributions is a result of collisional evolution, the significant difference between the faint-end slopes poses a challenge. One possible explanation would be that just as different accretion environments can lead to different bright-end slopes, non-uniform collisional dynamics in the primordial trans-Neptunian disk could have resulted in the color populations experiencing different early collisional histories owing to their different formation regions. Various areas of the primordial disk may have been characterized by a wide range of impact velocities and intrinsic collision probabilities.  In such a model, the currently-observed discrepancy between the faint-end slopes would be a relic of the pre-emplacement collisional evolution of the two color populations. Indeed, very little is known about the nature of the early Solar System, so one could not exclude this possibility. 

That said, the fact that the overall Trojan population is characterized by a faint-end slope so close to the canonical collisional equilibrium slope suggests that perhaps there is another explanation in which the two color populations experienced a similar collisional evolution within the primordial trans-Neptunian disk and were emplaced with similar faint-end slopes. In this case, the different faint-slopes would be explained by positing a mechanism that converts R objects to LR objects, hence flattening the faint-end slope of the R population, while simultaneously steepening the faint-end slope of the LR population. 

Previous laboratory work has shown that irradiation of surfaces rich in terrestrial bitumens and other organic compounds, which tend to have a characteristic red color, leads to the flattening of the spectral slope and a resulting less red color \citep{moroz,kanuchova}. However, since the incident radiation flux on the surface of a spherical body scales in tandem with size, this flattening effect is expected to be the same across the full range of Trojan sizes and hence does not explain the discrepant faint-end slopes observed in the magnitude distributions of the color populations. Furthermore, the timescale for flattening the spectrum of a R Trojan is much smaller than the time that has elapsed since emplacement and formation \citep{melita09}, so if irradiation is the sole mechanism for converting R objects to LR ones, one would not expect any R objects to remain. In \cite{melita09}, an additional mechanism is proposed whereby minor cratering events disrupt the spectrally flattened irradiation crust and excavate underlying material, which the authors of that work posit as being red in color, consistent with that of typical surfaces rich in complex organic materials. The added contribution of cratering leads to irradiated LR objects becoming R objects once again through resurfacing, thereby preventing all the R objects from turning into LR objects. However, the characteristic collisional timescale and, correspondingly, the timescale of resurfacing decrease with decreasing asteroid size, while the rate of irradiation is the same for all objects, as mentioned earlier. Therefore, the resurfacing of Trojans through cratering becomes more effective at returning irradiated LR objects to R objects when one goes to smaller sizes. This would lead to a relative excess of R objects at faint magnitudes, which is the opposite of what is evident in the observed color distributions. 

In this work, we suggest an alternative explanation for the discrepancy in faint-end slopes and examine the possibility that the fragments resulting from a catastrophic shattering impact on a R object are LR. In other words, we hypothesize that R and LR Trojans have more or less identical interiors, differing only in the spectroscopic properties of their outer surfaces, and that the destruction of red objects is the primary mechanism by which R objects become LR, thereby resulting in a relative depletion of red Trojans in the range of sizes for which shattering collisions have been significant. To assess the viability of this conversion hypothesis, we ran a series of simple numerical simulations that model the collisional evolution of the Trojan population since emplacement. The mechanics of our algorithm are similar to those used in previous studies of Trojan collisions \citep[c.f.,][]{marzari}. Earlier works have shown that the overall Trojan$-$Trojan collisional frequency among large objects with $H > 9$ is very low ($\ll$ 1 Gyr$^{-1}$). This means that most of the collisional activity is concentrated in the faint-end of the magnitude range, and that the magnitude distribution of bright objects is expected to remain almost unchanged over the age of the Solar System. Therefore, we only considered initial magnitude distributions that are broken power-laws of the form described in equation~\eqref{distribution} with a bright-end distribution identical to that of the currently-observed population ($\alpha_{1} = 1.11$, $H_{0} = 7.09$, and $H_{b} = 8.16$). For the initial faint-end slope, we considered values ranging from 0.45 to 0.55, in increments of 0.01. Objects in the initial population with absolute magnitudes in the range $H=7\rightarrow 23$ were divided into 50 logarithmic diameter bins using the conversion formula $D = 1329\times 10^{-H/5}/\sqrt{p_{v}}$, where we have assumed a uniform geometric albedo of $p_{v} = 0.04$ \citep{fernandez}. 

The initial color populations were constructed by taking constant fractions of the total population across all bins; based on the calculated 0.1~mag bin number ratios between R and LR Trojans in the bright-end portion of our data, we considered initial R-to-LR number ratios, $k$, ranging from 4 to 5, in increments of 0.5. The collisional evolution was carried out over 4~Gyr in 100,000 time steps of length $\Delta t= 40000$. At each time step, the expected number of collisions $N_{\mathrm{coll}}$ between bodies belonging to any pair of bins is given by
\begin{equation}N_{\mathrm{coll}}=\frac{1}{4}\langle P\rangle N_{\mathrm{tar}}N_{\mathrm{imp}}\Delta t (D_{\mathrm{tar}}+D_{\mathrm{imp}})^{2},\end{equation}
where $N_{\mathrm{tar}}$ and $N_{\mathrm{imp}}$ are the number of objects in a target bin with diameter $D_{\mathrm{tar}}$ and an impactor bin with diameter $D_{\mathrm{imp}}$, respectively; $\langle P\rangle = 7.35 \times 10^{-18}~\mathrm{yr}^{-1}~\mathrm{km}^{-2}$ is the intrinsic collision probability for Trojan$-$Trojan collisions and was approximated by the weighted average of the probabilities calculated by \cite{delloro} for L$_{4}$ and L$_{5}$ Trojans, taking into account the currently-observed number asymmetry between the two swarms. For a target bin with diameter $D_{\mathrm{tar}}$, only impactor bins with diameters satisfying the condition $D_{\mathrm{imp}} \ge D_{\mathrm{min}}$ were considered, where $D_{\mathrm{min}}$ is the minimum impactor diameter necessary for a shattering collision and defined as \citep{bottke}
\begin{equation}D_{\mathrm{min}}=\left(\frac{2Q_{D}^{*}}{V_{\mathrm{imp}}^{2}}\right)^{1/3}D_{\mathrm{tar}},\end{equation}
where $V_{\mathrm{imp}}=4.6~\mathrm{km}~\mathrm{s}^{-1}$ is the weighted average of the L$_{4}$ and L$_{5}$ impact velocities calculated by \cite{delloro}, and $Q_{D}^{*}$ is the strength of target. In our algorithm, we utilized a size-dependent strength scaling law based off one used by \cite{durda} in their treatment of collisions among small main-belt asteroids:
\begin{equation}\label{scaling}Q_{D}^{*} = c\cdot 10\cdot(155.9D^{-0.24} + 150.0D^{0.5} + 0.5D^{2.0})~\mathrm{J\,kg}^{-1},\end{equation}
where a parameter $c$ was included to adjust the overall scaling of the strength and varied in increments of 1 from 1 to 10 in our test trials.

\begin{figure}

\centering
        \subfigure[Comparison plot showing the final total distribution generated by the simulation (dashed black line) and the observed total distribution of Trojans, scaled to correct for incompleteness (white circles).]{
                \includegraphics[width=0.4\textwidth]{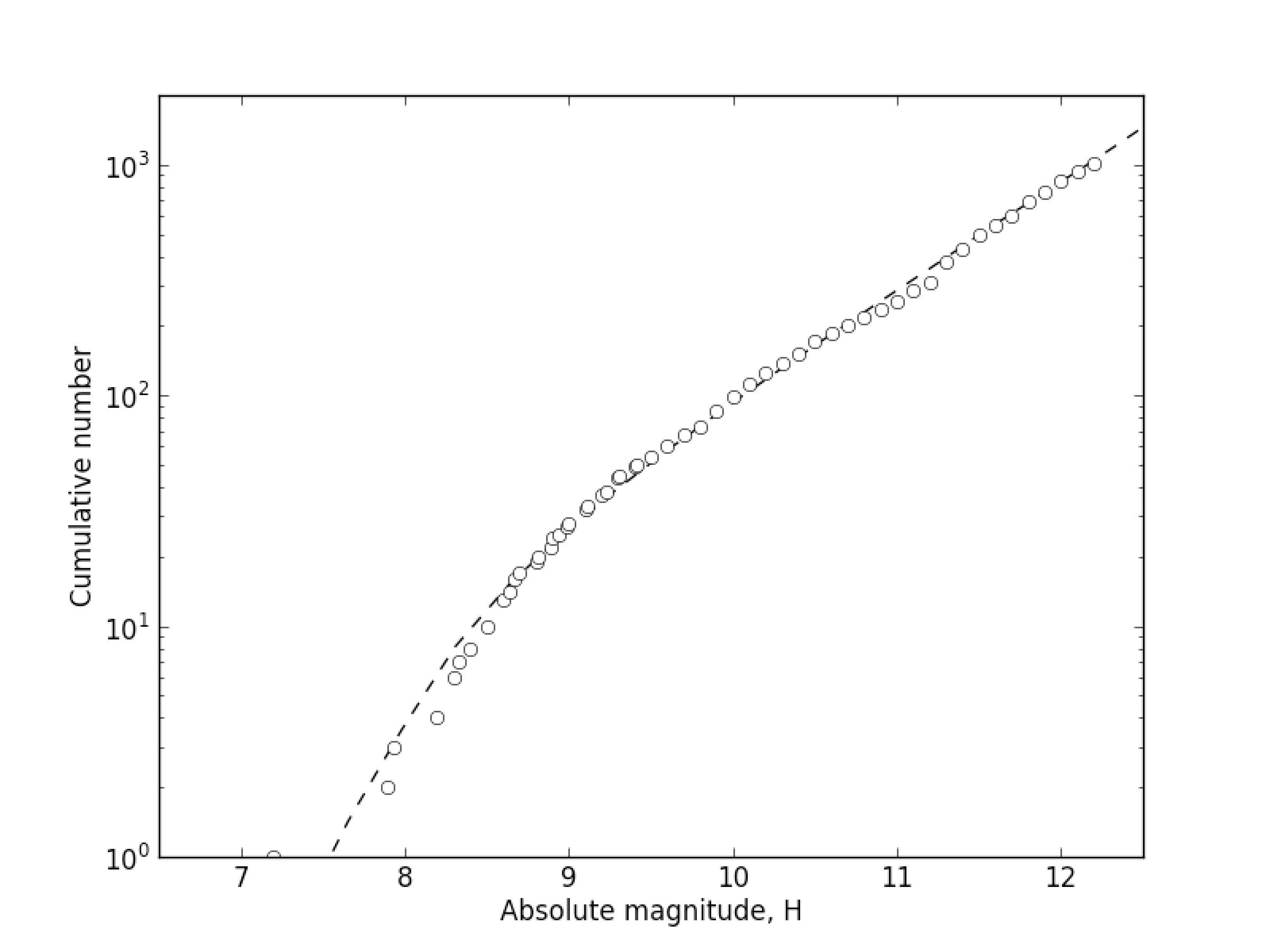}
            
                \label{fig:total}
       			}

        \subfigure[Same as (a), but for the R (red dashed line and squares) and LR (blue dash-dotted line and triangles) color distributions.]{
                \includegraphics[width=0.4\textwidth]{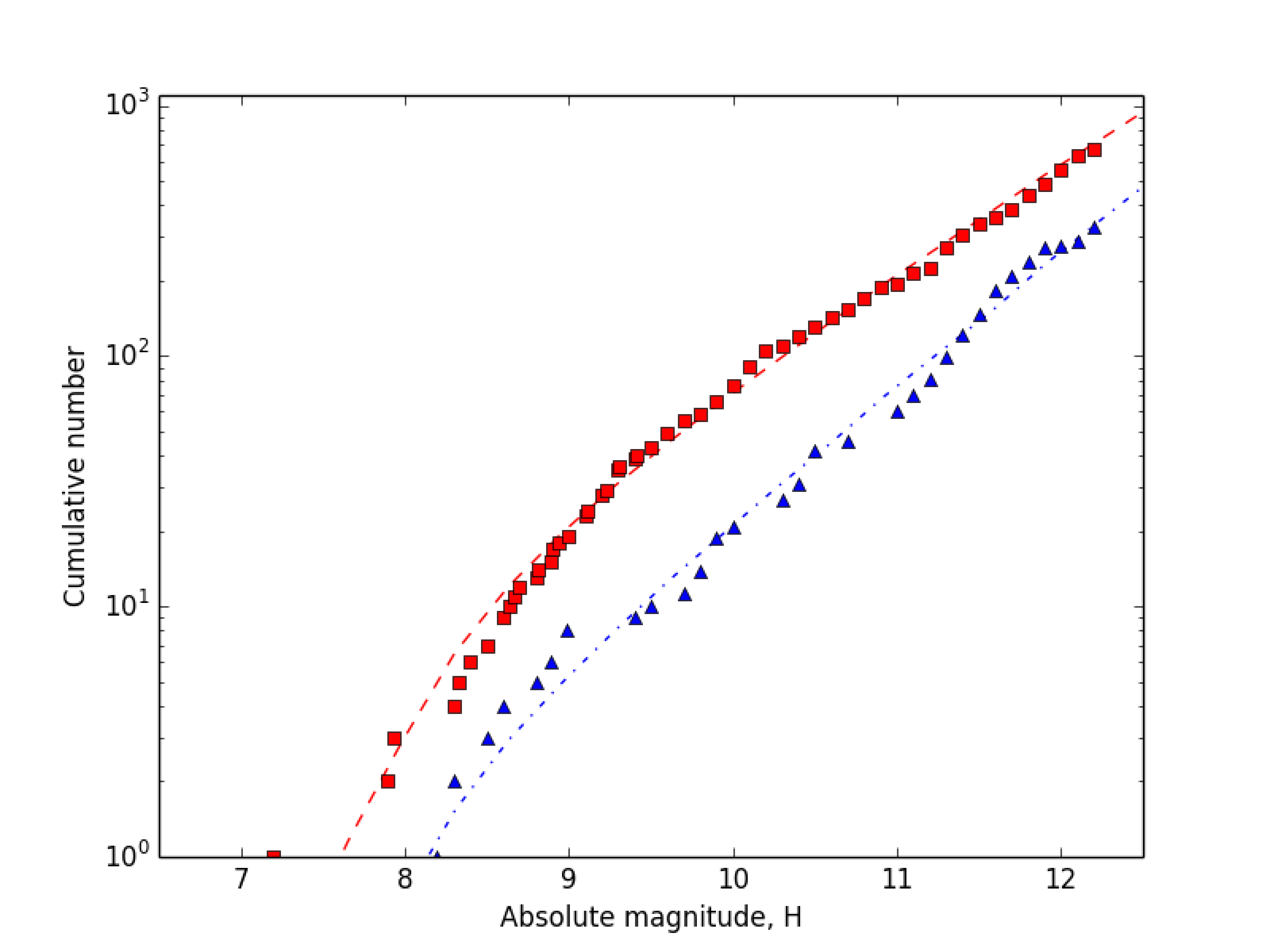}
               \label{fig:color}
      		}

        \subfigure[Plot comparing the best-fit distribution curves computed from the observed data (solid lines) with the final distribution curves generated by the simulation (dashed liens) for the total, R, and LR populations.]{
                \includegraphics[width=0.4\textwidth]{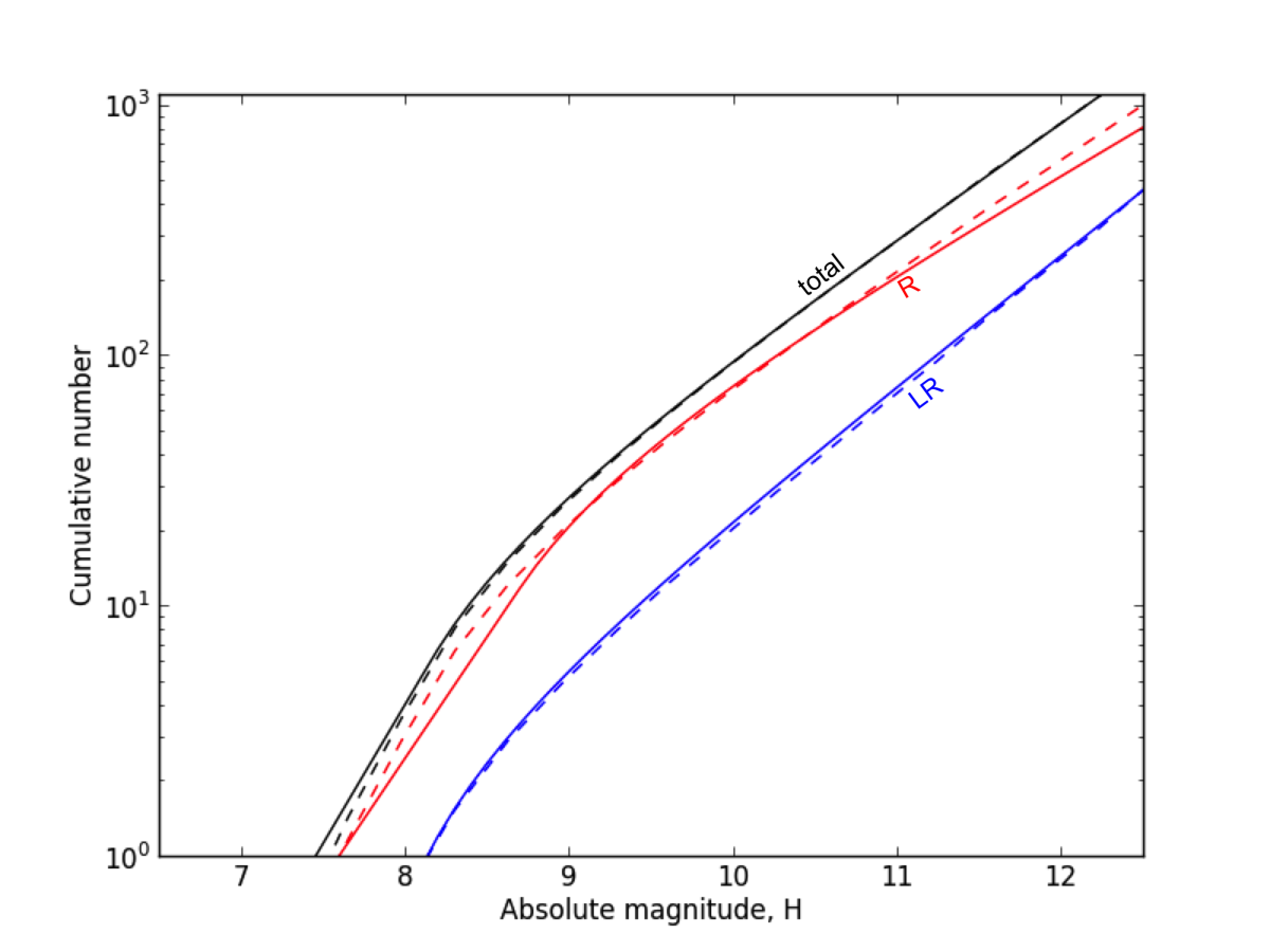}
        
                \label{fig:fits}
   			}

\caption{Comparison between the results from the best test run ($\alpha_{1}=1.11$, $\alpha_{2} = 0.47$, $H_{0}=7.09$, $H_{b} = 8.16$, $c = 6$, $k = 4.5$) and the observed Trojan magnitude distributions.}\label{fig:simulations}

\end{figure}

Our model tracked the collisional evolution of the two color populations separately and computed the number of collisions between objects of the same color, as well as collisions involving objects of different colors. For each time step, the number of collisions between all relevant pairs of bins was calculated, and the corresponding target and impactor numbers were subtracted from their respective bins. In all cases, regardless of the color of the target and/or impactor, the collisional fragments were redistributed into LR bins, thereby modeling the conversion of R objects to LR fragments through shattering. After running simulations for all possible values of the parameters ($\alpha_{2}$, $k$, $c$), we found that a large number of test runs yielded final total and color magnitude distributions that were consistent with the observed distributions analyzed in Section 3. To determine which run best reproduced the calculated faint-end slopes, we compared the simulation results directly with the fitted distribution curves. The test run that resulted in the best agreement with the data had an initial total distribution with faint-end slope $\alpha_{2}=0.47$, a strength scaling parameter $c = 6$, and began the collisional time integration with a R-to-LR bin number ratio $k = 4.5$. Plots comparing the final simulated distributions from this test run with the observed data are shown in Figure~\ref{fig:simulations}.

Although the simulations did not take into account other processes that may have affected the Trojan asteroids (e.g., dynamical dissipation), several conclusions about the evolution of Trojans can be made. First, the similarity between the initial test distributions that yielded good agreement with the data and the present-day total magnitude distribution indicates that collisional evolution has not played a major role in the post-emplacement development of the Trojan population, at least in the magnitude range we have considered in this work. In fact, our simulations are consistent with there being only 1 or 2 major collisions (involving asteroids with $D>100$~km) in the past 4~Gyr. To date, the only incontrovertible asteroid family that has been detected among the Trojans is the Eurybates family \citep{broz}, which shows that the currently-observed bright-end distribution is largely identical to the bright-end distribution of the primordial Trojan population at the time of emplacement. Second, the R-to-LR collisional conversion model has yielded simulated final color distributions that match the currently-observed color magnitude distributions well. This model is also supported by photometric data from members of the Eurybates family, all of which have very low spectral slope values that are consistent with LR objects \citep{fornasier}. Thus, the conversion hypothesis offers a feasible explanation for the curious faint-end slope discrepancy between the R and LR populations. 

The R-to-LR conversion model assessed here is attractive because it has some basis in recent work on KBOs, which, in the Nice model, arise from the same body of material as the Jupiter Trojans. The Kuiper belt is comprised of several sub-populations, among which are the so-called ``red'' and ``very red'' small KBOs \citep{fraser,peixinho}. A recent hypothesis describes a scenario in which KBOs formed in the trans-Neptunian disk at a range of heliocentric distances \citep{brown}. During formation in the primordial disk, all of these objects would have accumulated a mix of rock and volatile ices of roughly cometary composition. After the disk dissipated, the surface ices on these bodies began sublimating from solar radiation, leading to differential sublimation of individual ice species based on the location of the object. Whether a particular volatile ice species on the surface of these objects is retained or sublimates away is dependent on the volatility of the ice species and the temperature of the region where the object resides. As a result, for each ice species, there would have existed some threshold heliocentric distance for which objects at greater heliocentric distances would have retained that ice species on their surfaces, while those that formed closer in would have surfaces that were completely depleted in that ice species.  Irradiation of surface ices would lead to significant darkened irradiation mantle, which serves to protect ices embedded deeper down from sublimation and the further action of irradiation. Therefore, the hypothesis in \citet{brown} argues that the presence or absence of one particular volatile ice species may be the key factor in producing the observed bimodality in color among the small KBOs: Objects that retained that volatile ice species on their surfaces formed a ``very red'' irradiation mantle, while those that lost that volatile ice species from their surfaces formed a ``red'' irradiation mantle.

If the LR and R Jupiter Trojans were drawn from the same two sources as the ``red'' and ``very red" KBOs, any exposed volatile ices on the surface would have evaporated away during the process of emplacement to smaller heliocentric distances. In our hypothesis, we posit that the more intense irradiation at $\sim 5$~AU flattens the spectral slope of the irradiation mantles that formed prior to emplacement. As a result, the Trojans that formed a ``red'' irradiation mantle would be left with surfaces that appear relatively less red, while those that formed at greater heliocentric distances and developed a ``very red'' irradiation mantle would end up with the relatively redder surfaces characteristic of R Trojans. When a Trojan shatters during a catastrophic impact, the irradiation mantle on the surface would disintegrate and any newly-exposed volatile ices in the interior (including, crucially, the particular species responsible for the formation of the ``very red'' irradiation mantle) would sublimate away within a relatively short timescale. Thus, if one assumes that LR and R Trojans have similar interior compositions, the fragments resulting from the shattering of a R Trojan would indeed be spectroscopically identical to those that would result from shattering a LR Trojan. Subsequent irradiation of these pristine fragments would eventually raise the spectral slope slightly, but not to the extent as would result if volatile ices were retained on the surface. As a consequence, in the range of magnitudes for which collisions are significant, shattering events since emplacement would have gradually depleted the number of R Trojans while simultaneously enriching the number of LR Trojans.

Ultimately, the nature of the Trojans and the source of their bimodal color distribution may involve a complex interplay between several different physical processes. A full understanding of the origin of this color bimodality and the mechanisms that have shaped the Trojan color populations hinges upon better knowledge of the composition and chemistry of these objects, which may be obtained in the future with higher-quality spectroscopic observations.

\section{Conclusion}
In this paper, we have examined the magnitude distributions of the two color populations that make up the Jupiter Trojans. Earlier spectroscopic and photometric studies in the visible, near-infrared, and infrared were compared and shown to be consistent with one another, confirming the existence of two separate populations of Trojans whose members differ categorically with respect to various spectral properties. Using primarily spectral slope values calculated from the SDSS-MOC4 photometric data, we were able to categorize 221 Trojans with absolute magnitudes less than 12.3 into the R and LR color populations. In the process of compiling the data samples and evaluating for catalog and categorization incompleteness, we concluded that the current Trojan catalog is complete to $H=11.3$, while the SDSS is likely to have detected all Trojans in its coverage area with $H<12.3$.

Using the Kuiper$-$KS test, we demonstrated that the two color populations have magnitude distributions that are distinct to a high confidence level.  Fitting the distributions to a broken power law, we found that both the bright-end ($\alpha_{1}^{\mathrm R}=0.97^{+0.05}_{-0.04}$ versus $\alpha_{1}^{\mathrm LR}=1.25^{+0.09}_{-0.04}$) and the faint-end ($\alpha_{2}^{\mathrm R}=0.38\pm0.02$ versus $\alpha_{2}^{\mathrm LR}=0.52^{+0.03}_{-0.01}$) power-law slopes are different, with the most evident distinction in the faint-end portion of the magnitude distribution. Meanwhile, the total Trojan magnitude distribution is characterized by power-law slopes that are largely consistent with previously-published values ($\alpha_{1}=1.11\pm0.02$ and $\alpha_{2}=0.46\pm0.01$). The distinctness of the R and LR magnitude distributions suggests that the color populations likely formed in different regions of the primordial debris disk. The discrepancy between the faint-end slopes in particular may indicate that the color populations underwent different collisional evolutions before being emplaced into their current orbits. By running simulations of Trojan self-collisions, we have shown that this discrepancy is also consistent with a scenario in which the R objects differ from the LR objects only by the presence of a thin outer irradiation crust, and the color populations were emplaced with similar faint-end slopes. Subsequent shattering collisions could have led to the observed divergence of the faint-end slopes as all collisional fragments would be spectroscopically less red. Future study of Trojan asteroid spectra and composition promises to further our understanding of the origin and evolution of the two color populations.

\section*{Acknowledgements}
This work had its inception at the ``In Situ Science and
Instrumentation for Primitive Bodies'' study funded by the W.M. Keck
Institute for Space Studies. The authors were supported by NASA Grant NNX09AB49G. The authors also thank an anonymous reviewer for constructive comments that helped to improve the manuscript.

\end{document}